# Observations (from 2016 to 2020) of the Geminids from different regions of Russia by an amateur astronomer

*Filipp Romanov*[1]

I present the results of my observations (visual and photographic) of the Geminid meteor shower in 2016, 2018, 2019 and 2020. I observed meteors from different regions (Moscow and Primorsky Krai) of Russia, under different observation conditions: light pollution, Moon phases and weather. I used a DSLR camera with a lens to photograph meteor tracks. I compare the results of my visual observations in different years and determine the coordinates of the meteors from the photographs to graphically demonstrate the radiant.

## 1 Introduction

The Geminids are a prolific annual meteor shower and are favorable for observations in the Northern Hemisphere of the Earth. It is known that asteroid (3200) Phaethon is the parent body of this meteor shower. Figure 1 shows the image of this asteroid taken (by my request) remotely using 0.355-m $f/6.2$ Schmidt-Cassegrain telescope of Abbey Ridge Observatory, Canada (Lane, 2018). The results of my astrometric measurements (for dates 2021 September 14 and 15: near aphelion) were published in the Minor Planet Electronic Circular MPEC 2021-S21 (Minor Planet Center, 2021).

I have observed meteor showers and submitted my observations to the VMDB: Visual Meteor Database (Roggemans, 1988) of the International Meteor Organization since 2013. I have used my camera Canon EOS 60D with 18–135 mm $f/3.5$–5.6 lens for photographing meteors during visual observations all these years (before when the display and camera lens were damaged in 2021).

I chose the Geminids to describe in this paper from all the meteor showers that I have observed, because I have observed a sufficient number (several hundred) meteors of this shower, from different regions of Russia under different sky conditions (but always in cold weather: at temperatures ranging from $-10$ to $-20°C$).

I tried to observe the Geminids for the first time from the science city of Korolyov (in Moscow Oblast) near maximum activity in December 2013 and 2014, but in this region, it is usually cloudy this month, therefore, I had to observe through gaps in the clouds for a short time, and on both occasions the observation time was no more than 10 minutes, and I saw only 2–3 meteors. In 2015 and 2017, there were no gaps in the clouds, and I could not observe the Geminids, so my only useful observation over those years was the observation in December 2016 from Moscow.

[1] Amateur astronomer, Russia. ORCID: 0000-0002-5268-7735
Email: filipp.romanov.27.04.1997@gmail.com

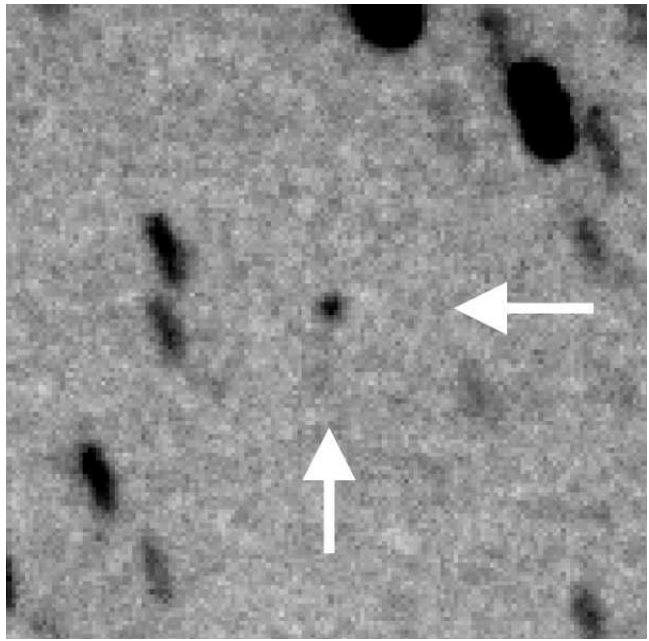

*Figure 1* – Stacked image of Phaethon from 10 photos (60 s exposure time, unfiltered) taken on 2021 September 15, from $05^h07^m$ to $05^h59^m$ UTC. North is up, field of view is 2 arcminutes.

## 2 My results of observations of Geminids for different years

### 2.1 Observations in 2016 from Moscow

In 2016, on the night of December 13/14, when I still lived in the room in the communal apartment in Moscow (geographic coordinates: $55°38'21.7''$ N, $37°40'23.3''$ E), I observed Geminids well for the first time. I monitored the weather, and after the sky cleared almost completely (which is very rare for December weather in Moscow), I went out to the loggia for observation, dressed warmly. The transparency of the atmosphere was poor (and made worse by the light of the full Moon and light pollution), I estimated the limiting magnitude (LM) at $+3.05$.

I observed and photographed from $21^h50^m$ to $22^h50^m$ UTC (from $00^h50^m$ to $01^h50^m$ by local time). During the observation, I saw 9 Geminid meteors: I estimated their brightness and recorded the direction and time of appearances (which I always do when observing meteors). My visual report was published in the VMDB under the number (ID) 74051.

Photographs were taken automatically in succession at focal length 18 mm, exposures of 11 and 12 seconds,

Table 1 – Data about meteors in 2016.

| No | Time (UTC) | Beginning $\alpha$ (°) | $\delta$ (°) | End $\alpha$ (°) | $\delta$ (°) |
|---|---|---|---|---|---|
| 1 | 21:53 | 91.81 | 1.51 | 89.69 | −2.12 |
| 2 | 22:02 | 113.42 | −18.61 | 113.45 | −22.91 |
| 3 | 22:05 | 104.24 | −10.06 | 103.47 | −13.77 |
| 4 | 22:34 | 103.75 | −22.94 | 103.05 | −26.93 |
| 5 | 22:37 | 114.79 | 7.83 | 114.79 | −3.75 |
| 6 | 22:47 | 90.72 | −5.69 | 87.47 | −11.54 |

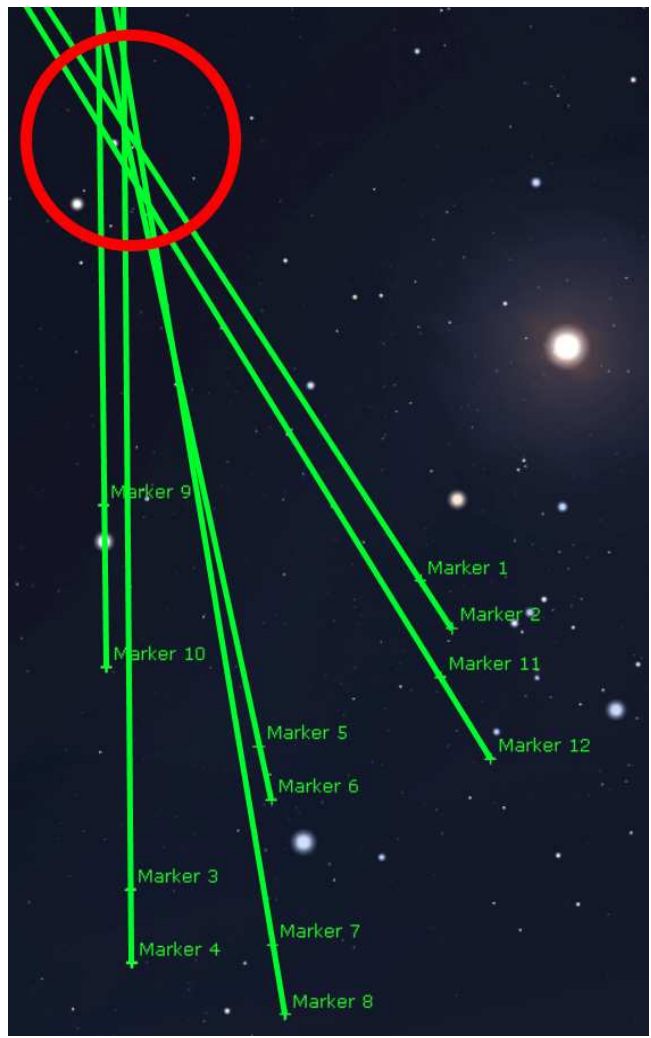

Figure 2 – Tracks of meteors in the sky chart according coordinates from photographs taken on the night of 2016 December 13/14.

ISO-250, $f/3.5$. Due to the blur of the stars in the photos during the exposure time, it is impossible to determine the positions of meteors precisely. Some meteors appeared in the interval between the opening and closing of the shutter (about 2 seconds of time), so they were not photographed. There are six meteors in the photos (of which I saw five visually).

I determined coordinates of the beginning and end of each of them (presented in Table 1), and put them in the form of markers on the gnomonic projection of sky map (in Figure 2) in Stellarium software. Then I drew lines through the points, and they approximately converge in the area of several degrees around the position: RA = 112.5°, Dec = +33.2° — near the Geminids radiant.

## 2.2 Observations in 2018 from Primorsky Krai

In 2018, I observed the Geminids from my small homeland: in Yuzhno-Morskoy (part of the city Nakhodka), Primorsky Krai (in the Far East). Geographic coordinates of the observation point: 42°51′30.3″ N, 132°41′17.6″ E. On the night of December 13/14 there was a haze in the sky and LM = 3 ... 4. Between $15^h30^m$ and $18^h00^m$ UT, I saw 12 Geminid meteors. My report ID = 78117 in the VMDB.

I observed it for a second time that year on the night of December 14/15. I saw 172 meteors of the Geminids (and 2 sporadic meteors) between $14^h30^m$ and $19^h16^m$ (with a few short breaks) UTC. It was a very beautiful and impressive show in the dark sky (LM = 4.5 ... 5.5). I often saw meteors and, in addition, comet 46P/Wirtanen (near the perihelion) was clearly visible (in the form of a nebulous star) with the naked eye.

My reports were published in the VMDB under the numbers: 78118, 78119, 78128, 78129, 78130, 78135, 78136, 78137. On those dates I also did free live streams of the Geminids on the Internet for viewers at my YouTube channel (the method was as follows: photographs taken with my DSLR camera were immediately shown on the computer and this image was shown live).

I made the composite photo from 14 photographs taken between $14^h06^m$ and $16^h42^m$ UTC using my DSLR camera (for each frame: exposure time 30 seconds, focal length 18 mm, ISO-1250, $f/3.5$). This was my first composite image of the Geminids (shown in Figure 3), the comet is also visible there. In total, several dozen meteors were recorded in the photographs taken on the night of 2018 December 14/15.

Based on visual observation data from several observers (including my observations) on the page "Geminids 2018 ZHR Graph – Peak" of the IMO website, a point with ZHR = 138.5 (error range: 128.84–148.24) has been determined for 2018 December 14, $16^h01^m$ UTC.

## 2.3 Observations in 2019 from Primorsky Krai

In 2019, the Geminids reached maximum just after full Moon. I observed (and made the live stream) this on the night of December 14/15 (also from Yuzhno-Morskoy).

Despite the bright moonlight, during the night (between $14^h20^m$ and $21^h30^m$ UTC with a few short breaks) I was able to see 64 meteors of the Geminids and 2 sporadic meteors; atmosphere transparency was good (LM = 4 ... 4.95).

I have found that the light of the Moon greatly reduces the number of visible meteors. IDs of my reports: 79739, 79750, 79767, 79768, 79769, 79770.

Figure 4 shows my composite image of meteors from 18 photographs (for each frame: exposure time 20 s, focal length 19 mm, ISO-800, $f/3.5$) taken from $14^h13^m$ UT to $16^h28^m$ UT.

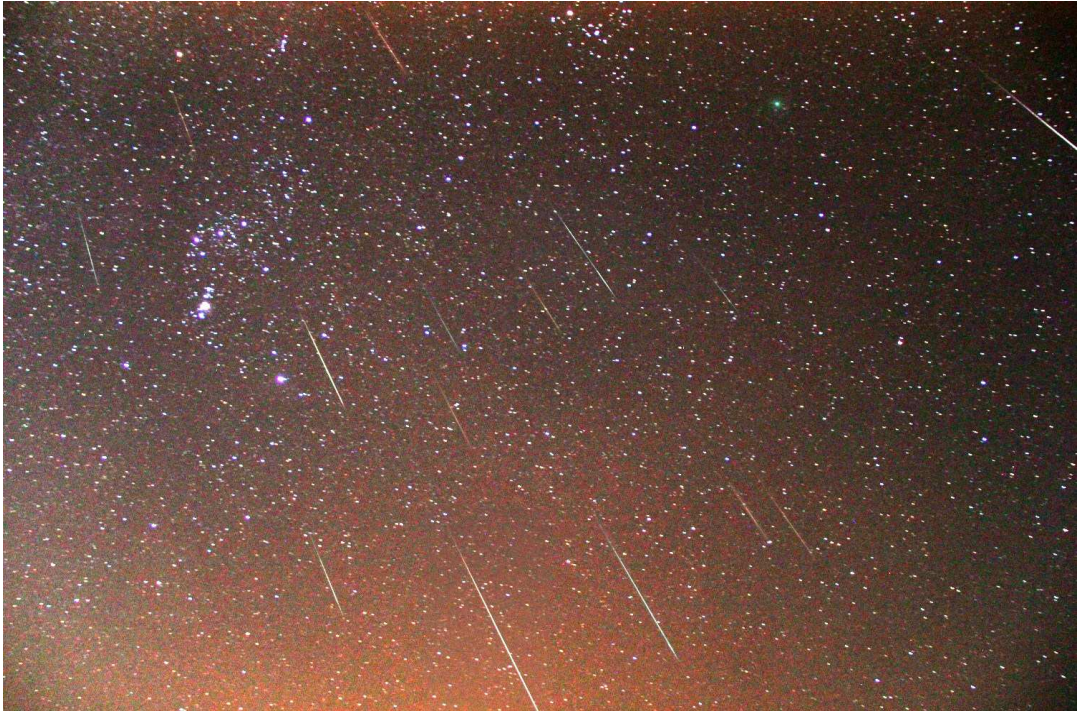

*Figure 3* – Composite image of the Geminids in 2018.

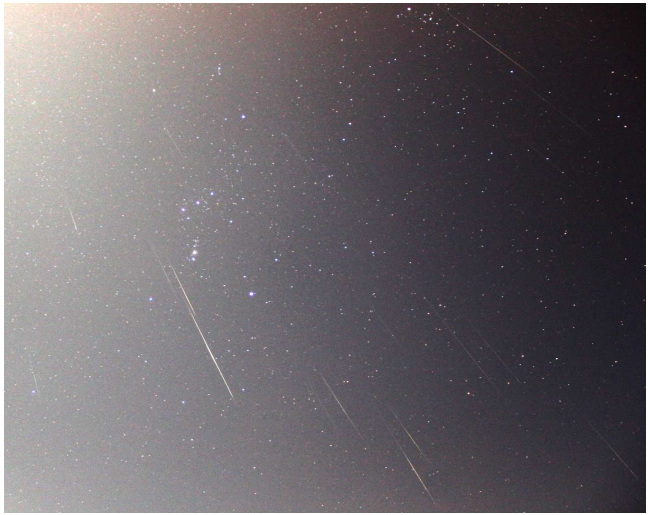

*Figure 4* – Composite image of the Geminids in 2019.

### 2.4 Observations in 2020 from Primorsky Krai

I observed (and did live stream) this meteor shower on the night of 2020 December 13/14, from Yuzhno-Morskoy. I saw 178 Geminid meteors (and one meteor from the Puppid-Velid meteor shower, and 16 sporadic meteors) between $15^h13^m$ and $21^h25^m$ (with a few short breaks) UTC. LM was 5.5 most of the time, but by morning it had become 4.5 ... 5.3.

I used a voice recorder to record information about the meteors I saw, including the magnitude and shower membership, and later I listened back to it and wrote down the data. IDs of my reports: 81864, 81924, 81926, 81927, 81928, 81929, 81940, 81941, 81942, 81943.

Figure 5 shows my composite image from 25 photographs (for each frame: exposure time 30 s, 19 mm, ISO-3200, $f/3.5$) taken from $14^h29^m$ UT to $18^h05^m$ UT.

*Table 2* – Data about meteors in 2020.

| No | Time (UTC) | Beginning $\alpha$ (°) | $\delta$ (°) | End $\alpha$ (°) | $\delta$ (°) |
|---|---|---|---|---|---|
| 1  | 14:16 | 57.80  | 16.13  | 55.10  | 14.53  |
| 2  | 14:29 | 73.04  | −3.59  | 68.06  | −8.94  |
| 3  | 15:15 | 85.55  | −6.88  | 84.55  | −8.43  |
| 4  | 16:02 | 66.44  | 15.84  | 62.28  | 13.52  |
| 5  | 16:03 | 86.93  | 11.80  | 83.90  | 8.80   |
| 6  | 16:07 | 90.26  | 21.71  | 86.35  | 19.38  |
| 7  | 16:12 | 83.45  | 5.19   | 82.03  | 3.57   |
| 8  | 16:15 | 96.91  | 9.48   | 94.35  | 5.35   |
| 9  | 16:17 | 111.80 | 2.44   | 111.48 | −3.38  |
| 10 | 16:26 | 110.30 | 10.11  | 109.70 | 5.86   |
| 11 | 16:28 | 91.43  | 26.33  | 88.99  | 25.40  |
| 12 | 16:43 | 115.45 | −7.35  | 115.83 | −16.59 |
| 13 | 16:43 | 102.47 | 21.81  | 100.97 | 20.10  |
| 14 | 16:45 | 72.53  | 5.75   | 68.00  | 2.06   |
| 15 | 16:52 | 97.01  | 12.94  | 94.84  | 9.89   |
| 16 | 16:52 | 95.46  | 6.37   | 92.91  | 2.10   |
| 17 | 16:54 | 107.50 | 16.75  | 106.48 | 14.11  |
| 18 | 16:57 | 117.36 | 12.79  | 117.80 | 9.87   |
| 19 | 17:02 | 88.02  | −9.15  | 85.95  | −12.61 |
| 20 | 17:20 | 119.23 | 17.00  | 120.31 | 13.04  |
| 21 | 17:20 | 132.11 | 10.93  | 134.69 | 7.65   |
| 22 | 17:26 | 125.13 | 12.71  | 126.56 | 9.63   |
| 23 | 17:27 | 123.77 | −7.31  | 124.59 | −11.51 |
| 24 | 17:31 | 128.78 | 16.02  | 130.57 | 13.81  |
| 25 | 17:35 | 104.79 | 17.82  | 103.63 | 15.62  |
| 26 | 17:37 | 125.17 | 10.04  | 127.06 | 5.49   |
| 27 | 17:38 | 113.64 | −6.10  | 113.65 | −9.00  |
| 28 | 17:43 | 103.10 | 7.17   | 101.72 | 3.51   |
| 29 | 17:58 | 90.54  | −4.80  | 88.11  | −8.97  |
| 30 | 18:04 | 112.21 | 23.07  | 111.50 | 19.68  |
| 31 | 18:05 | 115.96 | 26.41  | 116.12 | 25.58  |
| 32 | 18:05 | 88.33  | 6.87   | 83.93  | 1.97   |
| 33 | 18:08 | 133.48 | 21.81  | 136.61 | 19.55  |

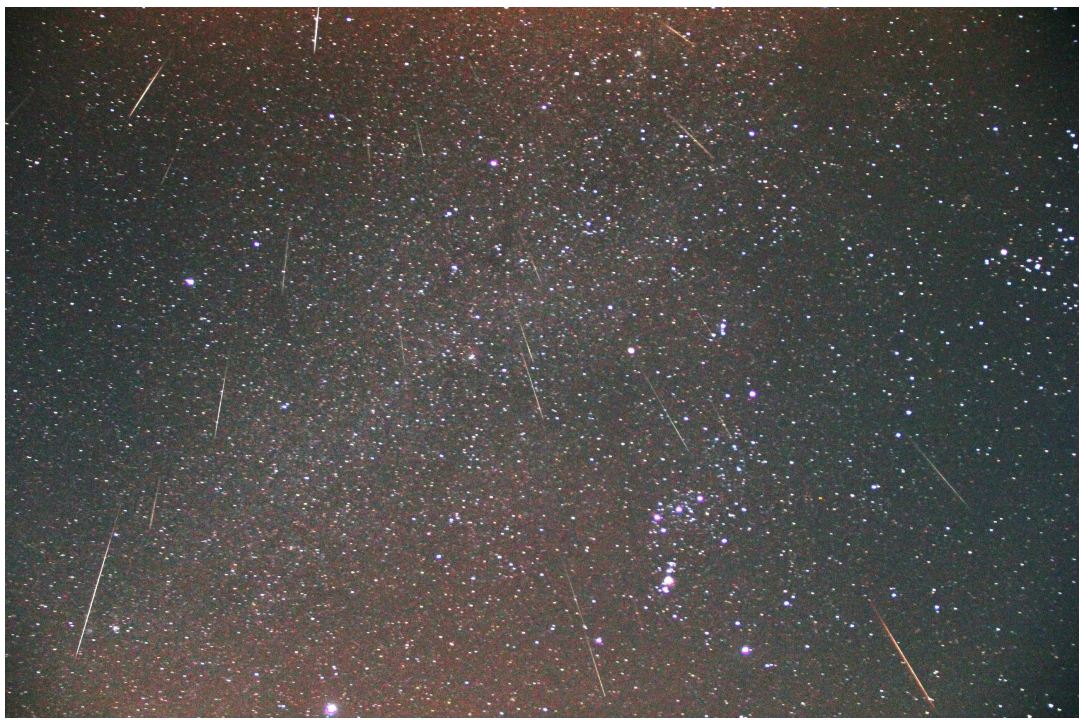

*Figure 5* – Composite image of the Geminids in 2020.

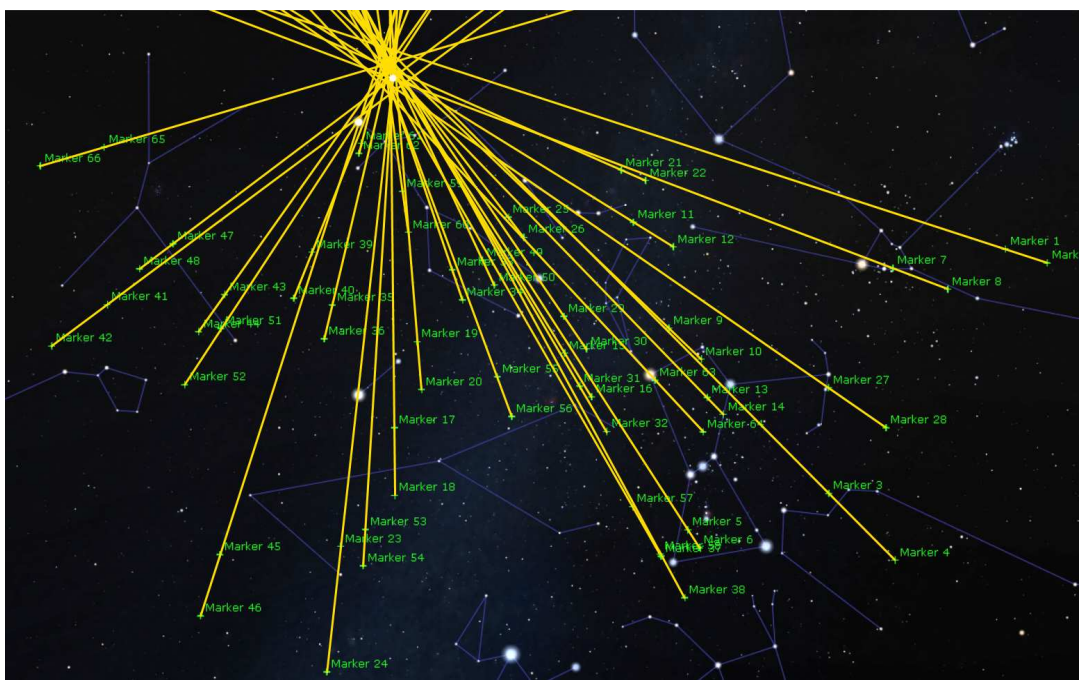

*Figure 6* – Tracks of 2020 meteors in the sky chart

I determined the coordinates (these data are presented in Table 2) of the beginning and of the end of each of 33 meteors (that were in my photos from $14^h16^m$ to $18^h08^m$ UT) and marked them in the sky map (gnomonic projection) in the Stellarium software. Taking into account the distortion at the edges of the lens, after continuation of the lines, they converge near the point RA = $113.5°$, Dec = $+32.5°$ (in the Figure 6): this almost coincides with the radiant of the Geminids. A total of about 50 meteors were captured in my photographs that night.

## 3 Conclusions

On 2017 January 23, I lost opportunity to live safely and unhindered in a room (and I lost access to my telescopes) in the communal apartment in Moscow, and I had to leave Moscow, but I continued to observe meteor showers from different regions of Russia, in which I was for some time during these years.

As a result of my visual observations of the Geminids over several years under different weather conditions and in different regions of Russia, I have concluded that in order to detect a large number of meteors of this shower, it is necessary to be in an area where the weather is often clear in December, and it is necessary to have the conditions: the absence of light pollution and the Moon below the horizon, and that the best time to observe is near the predicted maximum activity of this spectacular meteor shower. In my experience, due to the cold December nights during observations, it is necessary to wear warm clothes in order to prevent hypothermia of the body.

In the graphic way, I have shown that using even simple photographic equipment, it is possible to determine the approximate radiant of a meteor shower from the results of photography. To do this, one needs to determine the coordinates of the trails of meteors and plot them on the sky map in order to continue the lines.